**Title**

High-resolution measurements of face-to-face contact patterns in a primary school

**Authors**

Juliette Stehlé[1], Nicolas Voirin[2,3], Alain Barrat[1,4,*], Ciro Cattuto[4], Lorenzo Isella[4], Jean-François Pinton[5], Marco Quaggiotto[4], Wouter Van den Broeck[4], Corinne Régis[3], Bruno Lina[6,7] and Philippe Vanhems[2,3]

**Affiliations**

[1]Centre de Physique Théorique de Marseille, CNRS UMR 6207, Marseille, France

[2]Hospices Civils de Lyon, Hôpital Edouard Herriot, Service d'Hygiène, Epidémiologie et Prévention, Lyon, France

[3]Université de Lyon; Université Lyon 1; CNRS UMR 5558, Laboratoire de Biométrie et de Biologie Evolutive, Equipe Epidémiologie et Santé Publique, Lyon, France

[4]Data Science Laboratory, Institute for Scientific Interchange (ISI) Foundation, Torino, Italy

[5]Laboratoire de Physique de l'Ecole Normale Supérieure de Lyon, CNRS UMR 5672, Lyon, France

[6]Hospices Civils de Lyon, National Influenza Centre, Laboratory of Virology, Lyon, France

[7]VIRPATH, CNRS FRE 3011, UCBL, Université de Lyon, Faculté de Médecine RTH Laennec, 69372 Lyon, cedex 08, France

*Corresponding author : alain.barrat@cpt.univ-mrs.fr

**Abstract**

**Background**

Little quantitative information is available on the mixing patterns of children in school environments. Describing and understanding contacts between children at school would help quantify the transmission opportunities of respiratory infections and identify situations within schools where the risk of transmission is higher. We report on measurements carried out in a French school (6-12 years children), where we collected data on the time-resolved face-to-face proximity of children and teachers using a proximity-sensing infrastructure based on radio frequency identification devices.

**Methods and Findings**

Data on face-to-face interactions were collected on Thursday, October $1^{st}$ and Friday, October $2^{nd}$ 2009. We recorded 77,602 contact events between 242 individuals (232 children and 10 teachers). In this setting, each child has on average 323 contacts per day with 47 other children, leading to an average daily interaction time of 176 minutes. Most contacts are brief, but long contacts are also observed. Contacts occur mostly within each class, and each child spends on average three times more time in contact with classmates than with children of other classes. We describe the temporal evolution of the contact network and the trajectories followed by the children in the school, which constrain the contact patterns. We determine an exposure matrix aimed at informing mathematical models. This matrix exhibits a class and age structure which is very different from the homogeneous mixing hypothesis.

**Conclusions**

We report on important properties of the contact patterns between school children that are relevant for modeling the propagation of diseases and for evaluating control measures. We discuss public health implications related to the management of schools in case of epidemics and pandemics. Our results can help define a prioritization of control measures based on preventive measures, case isolation, classes and school closures, that could reduce the disruption to education during epidemics.

**Introduction**

The role of children in the community spread of respiratory infections such as influenza is a challenging epidemiological issue [1,2]. Children are thought to act as a reservoir as they are the first to be infected by various respiratory infections such as influenza, respiratory syncytial virus, or rhinovirus. They can be infected at school because of the numerous close contacts occurring between school children, and then act as sources of infection into their households from where infections can further spread in the community [1,2].

However, little is known about how children actually mix in a school environment [3-5]. An accurate description and understanding of the contacts between children at the school level would help to quantify the transmission opportunities of respiratory infections, and to identify the situations during school days where the risk of transmission is higher. Ultimately, one goal of such quantification would be to assess which control and containment measures have the best performance. In addition the availability of quantitative descriptions of the contact patterns between young individuals has the potential to inform mathematical models that aim at describing the propagation of diseases in populations.

In order to reduce this knowledge gap, the research priorities comprise collecting data on activities and interactions of children, in particular in schools. Until recently, most empirical studies have relied on self-reported information such as questionnaire-based surveys to determine mixing patterns [6-10]. Such surveys may however suffer from biases due to self-reporting. Recent advances in distributed sensing systems, based on mobile devices and wearable sensors, provide new ways of gathering data on human contacts and allow to mine the proximity relations and close-range interactions of individuals in real-world large-scale settings [3,11-17]. In this framework, a recent study [3] has for instance given important insights into the proximity patterns of students, teachers and staff in a US high school. As the data available to a broad research community remain scarce, additional measurement campaigns in different settings and covering different schools, countries, and age groups are much needed, in particular to test for common patterns and differences, and to understand their public health implications.

We deployed a proximity-sensing infrastructure based on radio-frequency identification devices (RFID) in a French primary school, and used it to collect, in an unsupervised manner [15], time-resolved data on the face-to-face proximity of children and teachers. We report the number and duration of contacts, and their dependence on age, class, daytime and school spatial structure. Based on the results, we highlight specific situations where children are in contact and during which infections may be transmitted, and discuss the potential impact of containment measures such as class or school closures during seasonal epidemics or pandemics of respiratory infections.

**Methods**

*Setting*

The study took place in a primary school in Lyon, France during two days in October 2009. The age of the students (elementary cycle, composed of 5 grades) ranges between 6 and 12 years. In this school, each of the 5 grades is divided in two classes, for a total of 10 classes. Each class has an assigned room and an assigned teacher. The smallest class has 22 children and the largest 26, for a total of 241 children and 10 teachers. 232 children and all teachers participated in the data collection. The school day runs from 8.30am to 4.30pm, with a lunch break from 12pm to 2pm, and two breaks of 20-25 min around 10.30am and 3.30pm. Lunches are served in a common canteen, and a shared playground is located outside the main building. As the playground and the canteen do not have enough capacity to host all the students at the same time, only two or three classes have breaks at the same time, and lunches are taken in two consecutive turns.

*Study design/Ethics statement and privacy*

The French national bodies responsible for ethics and privacy, the Comission Nationale de l'Informatique et des Libertés (CNIL, http://www.cnil.fr) and the "Comité de Protection des personnes" (http://www.cppsudest2.com/), were notified of the study, which was approved by the relevant academic authorities (by the 'directeur de l'enseignement catholique du diocese de Lyon', as the school in which the study took place is a private catholic school). In preparation for the study, parents and teachers were invited to a meeting in which the details and the aims of the study were illustrated. Verbal informed consent was then obtained from parents, teachers and from the director of the school. All participants were given a Radio-Frequency Identification (RFID) badge and were asked to wear it at all times. Special care was paid to the privacy and data protection aspects of the study: The communication between RFID badges, the readers, and the computer system used to collect data were fully encrypted. No personal information of participants was associated with the identifier of the corresponding RFID badge. The only piece of information associated with the unique identifier of the badge was the class the corresponding individual was associated with.

*Data collection infrastructure*

The measurement infrastructure, developed in the context of the SocioPatterns project [16], is based on active RFID devices, embedded in unobtrusive wearable badges. Detailed information on how this technology is used to monitor social interactions and to identify contact patterns is available in Refs. [17-19]. Individuals are asked to wear the devices on their chests, so that badges can exchange radio packets only when the individuals wearing them face each other at close range (about 1 to 1.5 m). This range was chosen as a proxy of a close-range encounter during which a communicable disease infection can be transmitted, for example, either by cough or sneeze, or directly by hands contact. The infrastructure parameters are tuned so that the proximity of two individuals wearing the RFID badges can be assessed with a probability in excess of 99% over an interval of 20 seconds [15,16]. This time scale allows an adequate description of person-to-person interactions that includes brief encounters. We define that a "contact" occurs between two individuals during an interval of 20s if and only if the RFID devices worn by the individuals exchanged at least one radio packet during that interval. After a contact is established, it is considered ongoing as long as the devices continue to exchange at least one such packet for every subsequent 20s interval. Conversely, a contact is considered broken if a 20s interval elapses with no exchange of radio packets. The detected proximity relations are relayed from the RFID devices to receivers installed in the monitored environment and eventually stored in a central system..

*Collected data*

Data on face-to-face interactions between 232 children (96% coverage) and 10 teachers (100% coverage) across 10 classes were collected over two days (Thursday, October 1$^{st}$ 2009 and Friday, October 2$^{nd}$ 2009, see Table 1). Data were collected from 8.45am to 5.20pm on the first day, and from 8.30am to 5.05pm on the second day. Contacts were not recorded outside of these time intervals. Radio receivers (RFID readers) covered all the classrooms, the canteen, the stairways, and the playground. No information on contacts taking place outside the school or during sports activities was gathered.

*Data analysis*

The patterns of contacts between children, and the corresponding mixing patterns between classes and age groups are analyzed through several quantities describing the number of contacts between individuals, the duration of these contacts, and the cumulated time spent in contact by each pair of individuals, as well as their statistical distributions characterized in particular by average and coefficient of variation squared ($CV^2$, defined as the squared ratio of the standard deviation to the mean of the distribution). A $CV^2 < 1$ indicates a distribution with low variance while a $CV^2 > 1$ indicates a high-variance distribution.

More precisely, we define the following weights quantifying the proximity relations of a pair of individuals *i* and *j*:

- the occurrence $o_{ij}$ is equal to 1 if at least one contact event between *i* and *j* was recorded during the measurement period, and 0 otherwise; this is a symmetric quantity ($o_{ji} = o_{ij}$);
- the frequency $n_{ij}$ gives the number of times that a contact event between *i* and *j* was recorded (hence $n_{ij} = 0$ if and only if $o_{ij} = 0$, and $n_{ji} = n_{ij}$);
- the cumulative duration $w_{ij}$ is the sum of the durations of the $n_{ij}$ contacts between *i* and *j* ($w_{ji} = w_{ij}$).

As children are grouped into classes, we also compute:

- the total number of contacts between children of classes A and B, $n_{AB} = \sum_{i \in A, j \in B} n_{ij}$ (for $A \neq B$; for $A = B$ we have $n_{AA} = \frac{1}{2} \sum_{i,j \in A} n_{ij}$), and the average number of contacts of a child of class A with children of class B, $n_{AB}/N_A$, where $N_A$ is the number of children in class A;

- the total time spent in contact between children of classes A and B $w_{AB}=\sum_{i\in A, j\in B} w_{ij}$ (for $A\neq B$; for $A=B$ we have $w_{AA}=\frac{1}{2}\sum_{i,j\in A} w_{ij}$), and the average contact time of a child of class A with children of class B, $w_{AB}/N_A$;
- the total number of contacts of children of class A, $\sum_B n_{AB}$, and the average number of contacts of a child of class A, $\frac{1}{N_A}\sum_B n_{AB}$ (where the sums on B include the case B=A);
- the total contact time of children of class A $\sum_B w_{AB}$, and the average time spent in contact by a child of class A, $\frac{1}{N_A}\sum_B w_{AB}$ (where the sums on B include the case B=A).

The quantities $n_{AB}$ and $w_{AB}$ define symmetric contact matrices at the class level, while the quantities $n_{AB}/N_A$ and $w_{AB}/N_A$ yield non-symmetric matrices taking into account the different class sizes [8]. All these quantities are computed both for the entire study duration and separately for each day.

In order to study the temporal structure of the interaction network, we also measure for each child $i$, and as a function of time, the number $k_i$ of other distinct children whom s/he was in contact with, as well as the total time $s_i$ the child $i$ spent in contact with other children since the beginning of the measurements. In other terms, $k_i$ is the degree of node i in the contact network aggregated since the start of the first day, and $s_i$ is the corresponding node strength, i.e., the sum of the weights of the links inciding on $i$. In each case, we distinguish between contacts with children in the same class ($k_i^{in}$, $s_i^{in}$) and with children of different classes ($k_i^{out}$, $s_i^{out}$). In terms of the proximity relations defined above, these quantities are therefore defined according to the following formulae (we consider in these formulae the example of a child $i$ of class A): $k_i=\sum_j o_{ij}$, $s_i=\sum_j w_{ij}$, $k_i^{in}=\sum_{j\in A} o_{ij}$, $s_i^{in}=\sum_{j\in A} w_{ij}$, $k_i^{out}=\sum_{j\notin A} o_{ij}$, $s_i^{out}=\sum_{j\notin A} w_{ij}$, where $o_{ij}$ and $w_{ij}$ are computed since the beginning of the first day. The temporal evolution of these quantities indicates how fast the overall contact patterns build up between the children, both at the class and at the school level.

We also build contact networks aggregated on a 20-minute timescale: each day is divided in sliding windows of 20 minutes, starting at intervals of 5 minutes and, for each 20-minute period, edges are drawn between those pairs of individuals for which at least one contact was recorded during this period. The average degree of each 20-minute network gives the average number of individuals with whom a given individual has been in contact with during the corresponding time window. By using these 20-minute sliding windows we filter out the fast fluctuations of the dynamical contact network and only retain the slowly-varying information on the network evolution.

As a summary of the contacts of each day, we additionally build two daily aggregated networks in which edges are drawn between a pair of individuals whenever at least one contact was recorded for that pair during the considered day. Each edge is weighted by the total time the corresponding individuals spent in contact during that day.

The two daily aggregated networks are compared using various measures. We compute the Pearson correlation coefficients between the characteristic parameters (number of contacts, total time spent in contact, etc.) measured for each individual in day 1 and in day 2. We also compare the network structures at a more detailed level, measuring the similarity between the neighborhoods of each node across the two days. A simple measure of similarity is given by the respective numbers of new and repeated distinct persons contacted in day 2 with respect to day 1. This can be further refined by specifying if these new and repeated contacts occur within the same class or with individuals of other classes. Moreover, as each link $i$-$j$ in the aggregated network is weighted by the total time $i$ and $j$ spent in face-to-face proximity (denoted by $w_{ij,1}$ for day 1 and by $w_{ij,2}$ for day 2), the similarity between the neighborhoods of an individual i in days 1 and 2 can be also quantified by the cosine similarity

$$sim(i) = \frac{\sum_j (w_{ij,1} w_{ij,2})}{\left(\sqrt{\sum_j w_{ij,1}^2} \sqrt{\sum_j w_{ij,2}^2}\right)}$$

This quantity is 1 if $i$ had contacts in both days with exactly the same individuals, spending the same fraction of time in proximity of each one, and 0 if, on the contrary, $i$ had contact in day 2 with totally different persons with respect to day 1.

Finally, by measuring the relative rates at which the RFID readers receive the packets emitted by individual badges, it is possible to perform approximate localization of the badges, and tell which RFID reader is closest to any given badge. Since the readers were installed in the classrooms, in the canteen, and in the courtyard, it is possible to detect in which of these areas each badge was situated at any point in time. This allows to construct the trajectories that children followed in space as they move within the school premises.

**Results**

*Number of contacts*

We recorded a total of 77,602 contact events involving 242 individuals (37,414 contacts on day 1 and 40,188 on day 2), with an average of about 317 contacts per individual on the first day (coefficient of variation squared $CV^2=0.22$) and 338 contacts per individual on the second day ($CV^2=0.27$).

Figure 1 reports the total number and cumulated duration of contacts involving children of a specific class or teachers. Figure 2 displays boxplots of the distributions of the individual contact numbers and cumulated durations, for each class and for each day. Figure 1 shows that the total number and duration of contacts involving teachers are smaller compared to those involving children, but Figure 2 indicates that this is mostly a consequence of the number of teachers being smaller than the number of children in a class. Both the number and the duration of contacts show a limited degree of heterogeneity across classes as well as across days. This is probably due to different class schedules (e.g., a class being absent during half a day because of sport activities) or to different school activities (e.g., group *vs* individual activities).

Each individual, on average, was in contact with 50 distinct individuals ($CV^2=0.14$) in day 1 and with 46.5 individuals ($CV^2=0.18$) in day 2. The corresponding distributions are shown as boxplots in Figure 2C, for each class and for each day.

*Duration of contacts*

Most contacts are of short duration, but contacts of very different durations are observed, including rather long ones. Figure 3 shows the distribution (number of events in each bin divided by the bin width) of the contact durations P(t), defined as the fraction of contacts with duration t. Figure 3 highlights that large fluctuations are observed in the contact durations. While the average duration of a contact is 33 seconds, and 88% of the contacts last less than 1 minute, more than 0.2% of the contacts exceed 5 minutes. Overall, no characteristic contact time scale can be defined (the squared coefficient of variation of the distribution is $CV^2=1.1$ ).

The heterogeneity of contact patterns is also observed for cumulated contact durations. Figure 3 displays the distribution of the total amount of time $w_{ij}$ that pairs of individuals *i* and *j* spent in contact during one day. Once again, most cumulated durations are short, but their distribution is broad: 64% of the pairs of individuals have interacted less than 2 minutes on the same day, 9% have spent more than 10 minutes together and 0.38% more than 1 hour. The average amount of time spent by two persons in face-to-face proximity during one day is 207 seconds (3min 27s) for day 1, and 236 seconds (3min 56s) for day 2, with squared coefficients of variation of the distributions of respectively 5.44 and 4.65.

When considering single individuals, the distribution of total time spent by an individual in face-to-face proximity with other individuals is more homogeneous, with an average of 10340 seconds (2h 52mn) for day 1 and 11000 seconds (3h 03mn) for day 2, with $CV^2=0.25$ for day 1 and $CV^2=0.33$ for day 2.

*Contact matrices*

Figures 4 and 5 display grayscale-coded matrices giving, at the intersection of row A and column B, respectively the total number of contacts ( $n_{AB}$ ) and the total duration of contacts ( $w_{AB}$ ) occurring between individuals of classes A and B during the two-day study. A clear hierarchical structure can be observed. Most contacts involve children of the same class, as shown by the whitish diagonal. Two-by-two light blocks around the diagonal also show that larger number and durations of contacts are observed between children of the same grade rather than with other grades. Finally, a separation between smaller grades (1st to 3rd) and upper grades (4th and 5th) grades is also apparent, most probably due to the lunch break schedule. The numerical values corresponding to Figures 4 and 5 are given in Tables 2 and 3, while Table 4 reports the average daily numbers and durations of contacts of an individual of one grade with individuals of another grade. Values are higher on the diagonal, the separation between smaller and upper grades is apparent, and values tend to decrease when moving away from the diagonal. Moreover, even when only contacts between children of different classes are taken into account, Tables 2 and 3 show that each class has more contacts with the other class of the same grade than with any other class, as also reported in [20]. Finally, teachers have sparse contacts with one another because they spend most of the time in class with children.

*Temporal evolution of the contact patterns*

Figure 6 shows the time evolution of the 20-minute aggregated networks for each day of the study. The number of individuals is stable during teaching hours, morning and afternoon breaks, and drops during the lunch time as some children were going back home to have lunch (the school is located in an urban area and many children actually live nearby and can go home for lunch). The average degree of the network displays a more interesting behavior, as it peaks at various moments, each corresponding to a break or the beginning or end of the lunch of a series of classes.

Figure 7 displays the time evolution of the quantities $k_i$, $k_i^{in}$ and $k_i^{out}$ averaged over all children. The average number of distinct persons contacted grows initially rapidly, mostly because of contacts occurring within each class. The average $k_i^{in}$ however saturates at the average class size after a few hours, meaning that each child has been in contact with all members of his/her own class, while new contacts across classes occur only during the breaks, leading to plateaus in the evolution of the cumulated average degree. In the second day, contacts within each class are the same as in the first day, and the average of $k_i$ continues to evolve only during the breaks due to contacts involving children of different classes that had not occurred on the first day.

Figure 8 gives more insight into the evolution of the contacts of the children by taking into account the cumulated time spent in contact. It shows the time evolution of $s_i$, $s_i^{in}$ and $s_i^{out}$, averaged over all children. The average contact time spent by a child with other children grows regularly with time, in a similar way in both days. While the time spent with other children of the same class also has a regular increase (only slightly faster during morning and afternoon breaks), the time spent with children of a different class evolves significantly only during the lunch break (the evolution occurring during the morning and afternoon breaks are much smaller). Overall, at the end of each day, a child has spent on average three times more time in face-to-face proximity with children of his/her class than with children of other classes.

Video S2 presented as Supplementary Information gives a visual summary of the contact patterns occurring during the first day. It highlights the existence of many timescales in the contact patterns, with both short and long lasting edges. It also illustrates the periods in which edges exist mostly within each class, contrasted with periods such as breaks, during which mixing between classes takes place. In the latter case, it allows to understand which classes have most frequent contacts with one another.

*Daily aggregated networks*

Figure 9 displays the aggregated contact network for the first day of the study. For ease of interpretation, edges between individuals who spent together a cumulated time smaller than 2 minutes during the day have been removed. This corresponds to keeping only the strongest 33.2% of all edges. The figure highlights the mixing patterns between children of different classes and how children preferentially mix within the same class or age group. Classes within the same grade tend indeed to be more connected than classes belonging to different grades.

*Comparison between day 1 and day 2*

A comparison between the characteristics of the overall face-to-face contact patterns in the two days of the deployment is reported in Table 5. Statistical quantities such as the average total number and durations of contacts, the number of different persons contacted, or the contact durations are extremely close across the two days. As shown in Figure 2 at the class level, the whole distributions are in fact similar.

At a more detailed level, the Pearson correlation coefficients between the number of contacts of an individual in the first and second day is 0.53; for the time spent in contact, it is 0.54; for the number of distinct persons contacted it is 0.53. These values show an overall strong correlation between the behavior of individuals from one day to the next.

Moreover, each child, on average, has 26 repeated contacts on the second day with children met during the first day (19 in the same class and 7 in a different class), and new contacts with 20 other children (1.4 in the same class, 18.4 in a different class). The average cosine similarity between his/her neighborhoods across the two days is 0.67 (0.74 for the neighborhood restricted to his/her own class, 0.2 for the neighborhood restricted to children in a different class). This indicates a repetitive pattern inside each class but a non negligible renewal of the contacts between classes across consecutive days.

*Trajectories in space*

Figure 10 displays the trajectories followed by children as they move across the classes and public spaces of the school. It shows how each class moves from its classroom to the courtyard and then comes back at various times. During the lunch break, some children go first to the cafeteria and then to the courtyard, encountering children who are moving in the opposite direction. It is apparent how these trajectories, dictated by the school schedule, strongly contribute to shaping the mixing patterns between classes and grades.

**Discussion**

To our knowledge, this is the first study presenting detailed measures of close (face-to-face) proximity interactions between children in a primary school (see however [3] for the case of a high school). These descriptive results on contact patterns are of interest for modeling the spread of various infectious diseases, and possibly for investigating the role of specific control measures, such as closure of classes, immunization strategies, and so on. Time-resolved contact data were collected over two school days by deploying a wireless sensor network of RFID badges that record close-range (1 to 1.5m) face-to-face proximity between individuals. The present study had a very high participation rate (> 95%). Relying on unobtrusive wearable devices allows the unsupervised detection of contacts during which a communicable disease, in particular a respiratory disease, may be transmitted. This is an important advantage compared to approaches based on questionnaires, especially among the youngest.

*Comparison with previous studies*

A number of other studies describe or estimate social contact numbers and durations [3-10]. Comparison with previous results is clearly important but is made difficult by differences in the definitions of interaction/contact as well as by differences in the measurement techniques. As the present study considers the unsupervised detection of face-to-face proximity, it does not rely on surveys nor on the memories of participants. It is thus expected that larger total number and durations of contacts will be obtained, in comparison with survey-based methods.

Table S1 reports the comparison of the number and duration of contacts between previous studies and the present one. As expected, when all contacts are taken into account, we obtain larger values than the studies cited above, with the exception of [3]: as the infrastructure described in [3] considers a broader detection range (3 meters proximity) than in the present case, it is not surprising that our study detects less numerous and shorter contacts. We report that each child has on average 323 contacts lasting 33 seconds per day with other children, corresponding to contacts with an average of 47 distinct other children, for an average daily total interaction time of 176 minutes. The present study gives however access to much more detailed information such as the duration of each contact and the cumulative duration of the contacts between two individuals. An important result concerns the fact that most contacts are short but that the distribution of contact durations is very heterogeneous, with a non-negligible fraction of long-lasting contacts. Strongly heterogeneous distributions of contact durations such as the one displayed in Figure 3 have been observed in other settings, including conferences involving only adults [15,17] and a US high school [3]. Observing similar patterns among young children was however not a priori expected.

To allow a more informed comparison between studies based on different methodologies, we compute for each child or for each pair of individuals the number and total duration of contacts lasting longer than a given threshold. The results are summarized in Table 6. For instance, when restricting to cumulated contact durations of at least one minute, the number of different children with whom a child has interacted drops to 21, and the corresponding total interaction time drop to 163 minutes. Moreover, numbers close to those reported by Glass *et al.* [9], Zagheni *et al.* [6] and Del Valle *et al.* [21] are obtained when one takes into account only pairs of children having interacted for at least 10-12 minutes per day. Overall, our results are therefore quantitatively different from other studies, as can be expected from the strong methodological differences, but become compatible with previous studies when applying filtering procedures which retain only the longest contacts.

*Public health implications in the field of infectious diseases*

Our results show that children mix preferentially with children within their age group. This effect, known as age homophily, is largely due to the fact that children study together and have the same schedule, and represents a general feature studied in various contexts by sociologists [22]. As a result, the contact matrices display a hierarchical block-diagonal structure once their entries (classes) are sorted according to the seniority of children, as visible in Figures 4 and 5 (Tables 2 and 3): in addition to the strong diagonal entries (which correspond to the clusters of Figure 9, i.e., to contacts within classes) one can see blocks corresponding to the same grade (e.g., 1A/1B, 2A/2B, 3A/3B, etc.) and two larger blocks that span the junior grades (1 to 3) and the senior grades (4 to 5). In addition, the study of the daily evolution of the contact patterns allows to detect periods of higher contact activity, corresponding to breaks or lunch time, as well as moments when the contact weights (i.e., the cumulative time in contact) increase the most.

These results may help to advise public decision-makers on interventions aimed at containing or mitigating the propagation of communicable diseases at the level of schools, in particular in case of an epidemic or a pandemic. School closure has been proposed as an effective physical intervention to reduce transmission of respiratory pathogens, especially influenza [23]. However, it is not well understood how the benefit of closing entire schools, in terms of reducing cases, morbidity and mortality, compares to the economic costs of such interventions. In addition, the effectiveness of school closure depends on the effectiveness of other measures such as vaccination or antiviral drugs. Our results could be of interest in this context, especially if combined with other sources of information on the contact patterns of children [3,4,20]. The fact that a child spends three times more time in contact with classmates than with children of other classes suggests for instance that closing selected classes instead of the whole school could be a viable alternative. Additional intermediate steps between class and school closures could be devised through the analysis of aggregated contact networks, such as the one depicted in Figure 8, and exposure matrices such as the ones of Tables 2 and 3: classes most strongly linked to the class of the first detected case could for instance be closed in order to reduce the risk of propagation to the remaining classes. It would be interesting to assess by means of numerical simulation whether the closure of a single class or of a group of classes could efficiently mitigate the propagation of a disease at the school level. Finally, as highlighted by Figures 8 and 10 and by Video S2 of the Supplementary Information, preventive measures such as shifts of the class schedules could substantially reduce contacts between classes, which could be particularly relevant for preventing transmission events from asymptomatic cases.

*Informing mathematical models*

The development of mathematical models that aim at describing the spread of the infection and its prevention and control is hindered by the lack of information on the contact patterns between individuals. Epidemiological models of disease transmissions in structured populations depend heavily on the knowledge of the amount and duration of contacts between individuals of different age groups. To reduce this knowledge gap, we provide the exposure matrix of Tables 2 and 3. In these matrices, the cells at row *A* and column *B* give the average number and duration of contacts between one individual of category *A* with any individual of category *B*. This may help to refine the young age groups of the contact matrices proposed by Zagheni [6], Wallinga [7] and Mossong [8]. Given the important role of children and young adolescents in the community spread of respiratory infections, it is important to detail this part of the matrix. In particular, we remark that most contacts occur within a given class, and relatively few contacts occur across classes, effectively limiting the ability for diseases to spread between different classes.

These results highlight important properties of the contact patterns between school children that need to be taken into account when modeling the propagation of diseases and when evaluating control measures. On the one hand, our results tend to indicate that assumptions such as the homogeneity of contact durations, or a homogeneous mixing between classes, may yield misleading results. On the other hand, Figures 7 and 8 (together with video S2 of the Supplementary Information) show that homogeneous mixing within each class seems to be a good approximation. Of course, these considerations should be assessed by numerical simulations comparing the relative outcomes of different assumptions, such as global homogeneous mixing, homogeneous mixing within each class, and a network model.

*Caveats*

In the following, we discuss some limitations of the present study, and point to strategies for moving forward.

First of all, the deployed infrastructure only measured contacts between children while they were in the school building or in the playground. Badges were not worn during sport activities, which often involve close proximity situations and physical contacts. Moreover, even though the children would not be in school during a school closure, they would still mix with other children and adults in the community and spread the virus through these contacts. It would be interesting to use the data collection infrastructure to combine school data with household data and data on contact patterns during school closure [24,25]. Coupling the dynamical contacts patterns at school and at home would allow to improve our understanding of the role of children as a reservoir in the community spread of infections.

Another potential issue concerns the possibility that children changed their behaviour because they were wearing badges and knew they were participating in a scientific measure. According to observers familiar with the environment (teachers and staff), however, no significant change could be detected in the children's behavior, and the children seemed to rapidly forget about the badges. In addition, while detailed explanations were given to the parents about the study and the badges, details on the role of the RFID badges (e.g., their detection range) were not given to the children.

From a public health perspective, it has to be emphasized that the collected data provide information on the mutual proximity of badges (and therefore of the persons wearing the badges), but not on the occurrence of physical contacts. Our measurements may thus be used in the context of, e.g., respiratory-spread pathogens but not for infectious agents transmitted by skin contact. Note however that physical contact can only occur between persons who are already in spatial proximity. Therefore, it would be very interesting to study the fraction of close encounters that result in a physical contact. In the future, the use of devices that can directly sense physical contact (e.g., body-area networks) may be explored.

The short period of time (two days) of data collection also limits the ability to draw conclusions on what happens at longer time scales. Deployment of the sensing infrastructure over much longer timescales is needed in order to confirm the present results.

Finally, the data presented in this study depend on the school schedule and spatial structure, and the generalization of our results to other schools should be carried out with caution. Some properties are however expected to be rather general, such as the heterogeneity of the contact durations and of the cumulated contact durations, that has been observed in several other settings [3,11,15,18,19] and seems to be ubiquitous in human interactions.

*Perspectives*

Further research will use the gathered data to simulate the transmission of infectious agents (e.g. respiratory or gastro-enteric viruses) inside a school, to evaluate the role of the index case, and to assess the impact of various containment measures (e.g. class closure, homogeneous partial vaccination vs. vaccination of whole classes, at fixed coverage, etc). Further deployments in other schools with different schedules, other countries, and possibly for longer periods, will also be very useful to cross-validate our findings.

**Acknowledgments**

We warmly thank Bitmanufaktur, the Openbeacon project and Truelite for their technical support. We are particularly grateful to all children, their parents and the school staff who volunteered to participate in the data collection.

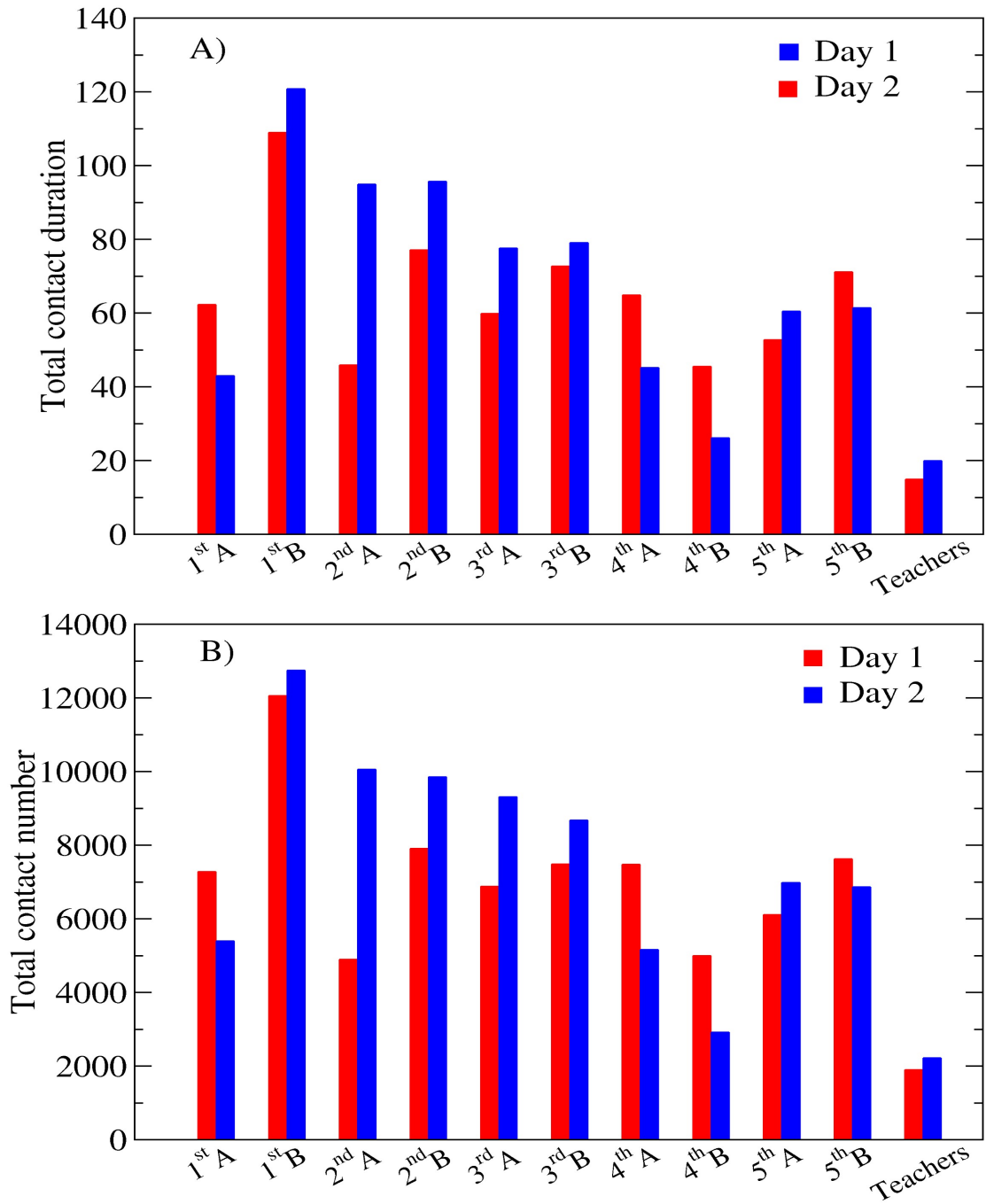

**Figure 1.** Total cumulated duration (A, in hours) and number of contacts (B) involving individuals of each class, for each day

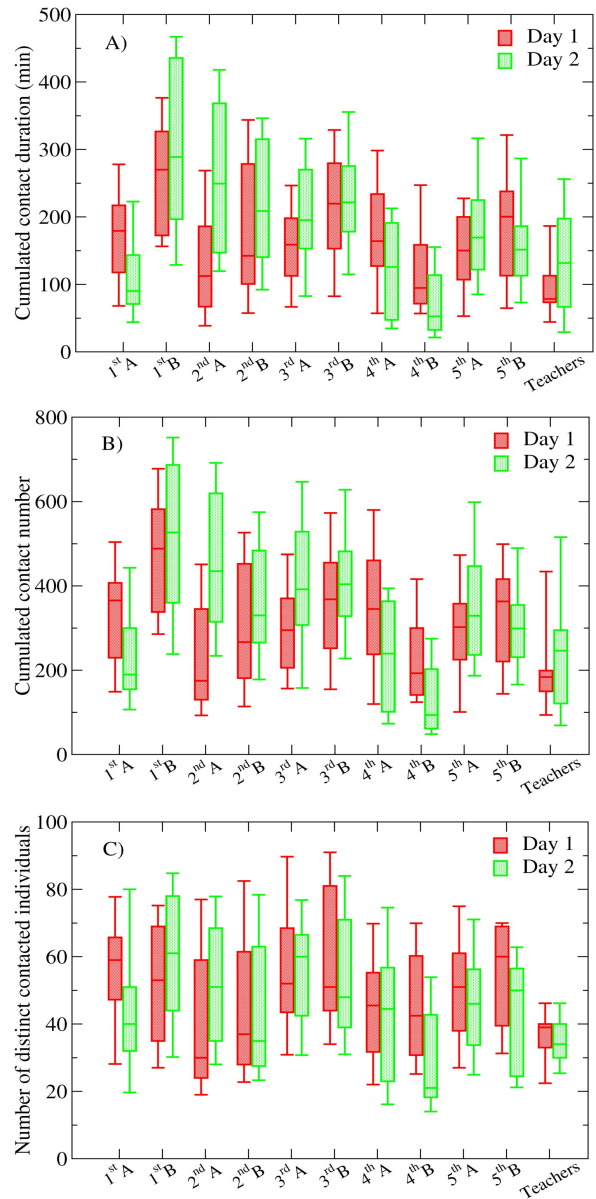

**Figure 2.** Boxplots of the distributions of the cumulated duration (A, in minutes) and number (B) of contacts involving an individual, for each class and for each day. Panel C gives the distributions of the number of distinct individuals with whom an individual of each class has had at least one contact. In each boxplot, the horizontal line gives the median, the box extremities are the 25[th] and 75[th] percentiles, and the whiskers correspond to the 5[th] and 95[th] percentiles.

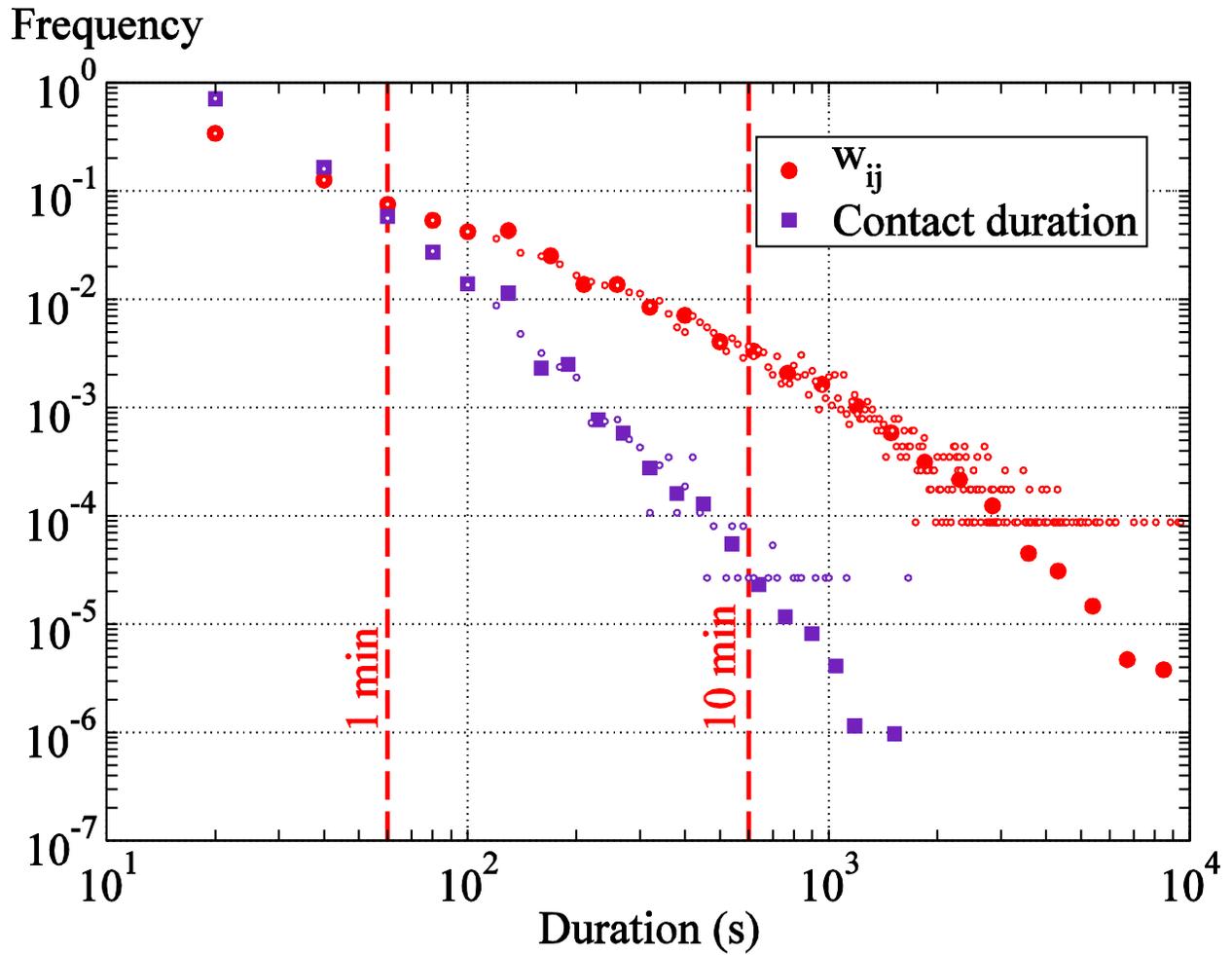

**Figure 3.** Log-log plot of the distribution of the contact durations and of the cumulated duration of all the contacts two individuals i and j have over a day ( $w_{ij}$ ). 88% of the contacts last less than 1 minute, but more than 0.2% last more than 5 minutes. For the cumulated durations, 64 % of the total duration of contacts between two individuals during one day last less than 2 minutes, but 9% last more than 10 minutes and 0.38% more than 1 hour. The small symbols correspond to the original distributions, and the large symbols to the log-binned distributions.

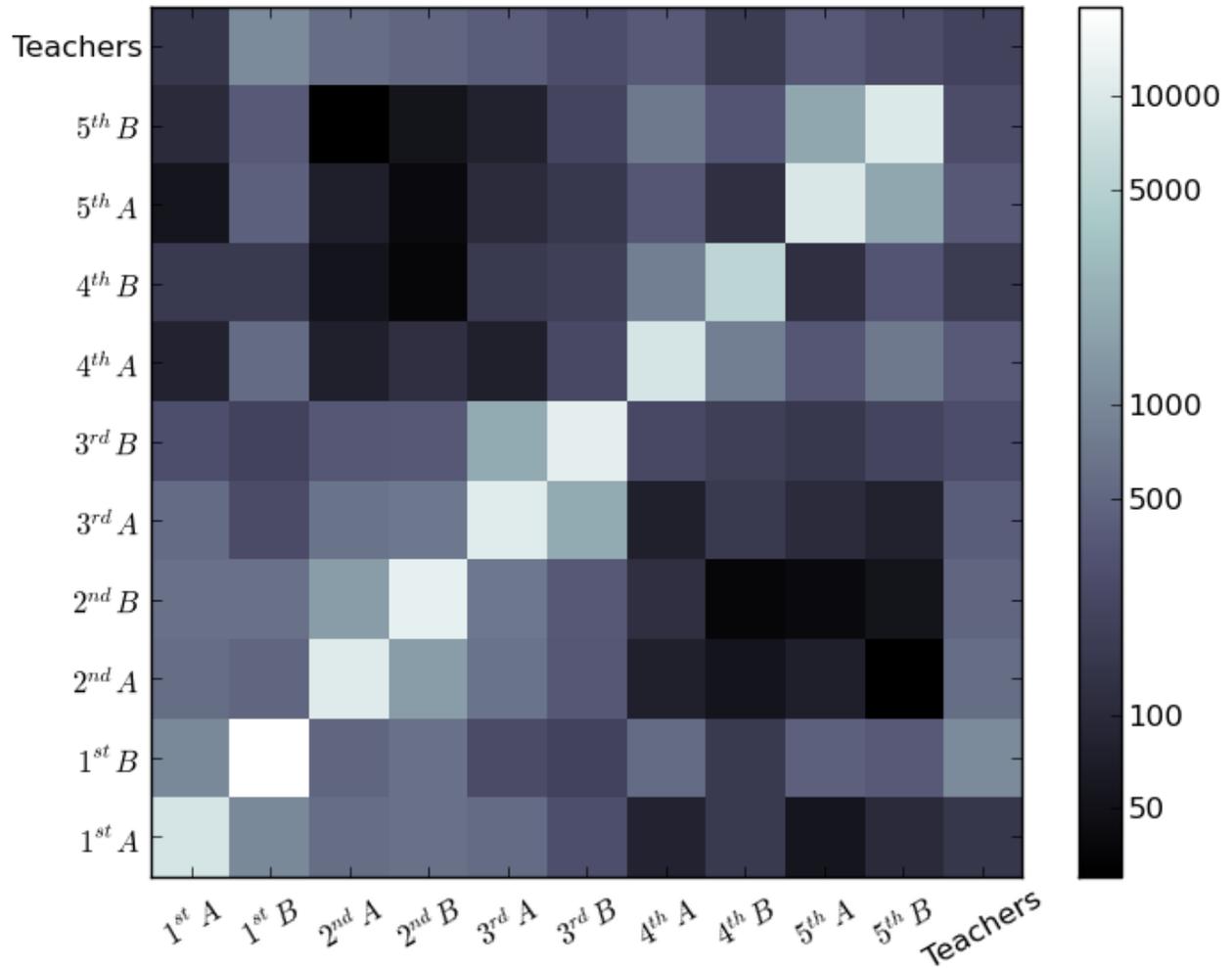

**Figure 4.** Grayscale-coded contact matrix between classes. The matrix entry for row A and column B gives the number of contacts ($n_{AB}$) measured between individuals of classes A and B over the two days of data collection. A logarithmic grayscale is used to compress the dynamic range of the matrix entries and enhance the off-diagonal hierarchical structure.

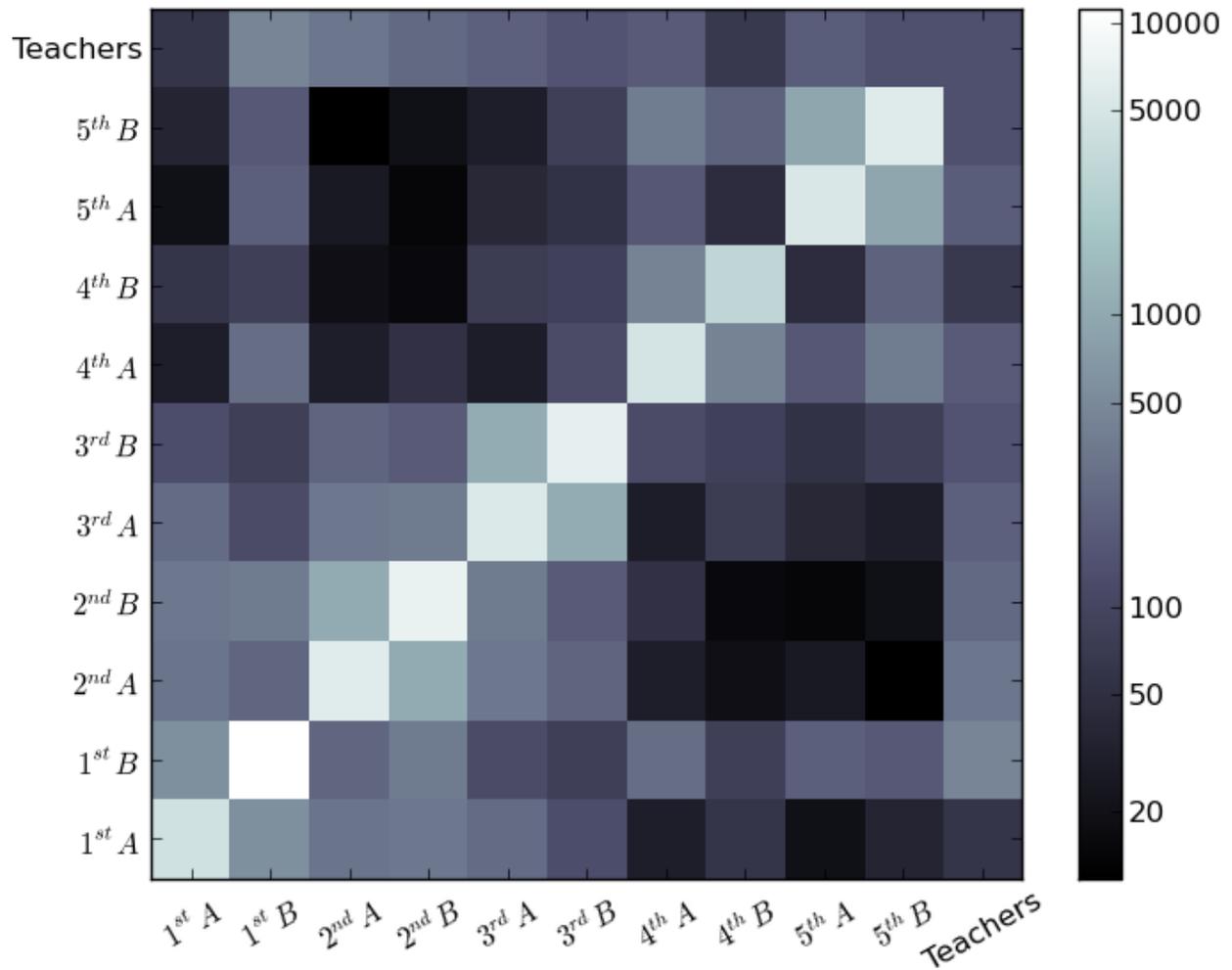

**Figure 5.** Grayscale-coded contact matrix between classes. The matrix entry for row A and column B gives the cumulated duration ($w_{AB}$, in minutes) of contacts measured between individuals of classes A and B over the two days of data gathering. A logarithmic grayscale is used to compress the dynamic range of the matrix entries and enhance the off-diagonal hierarchical structure.

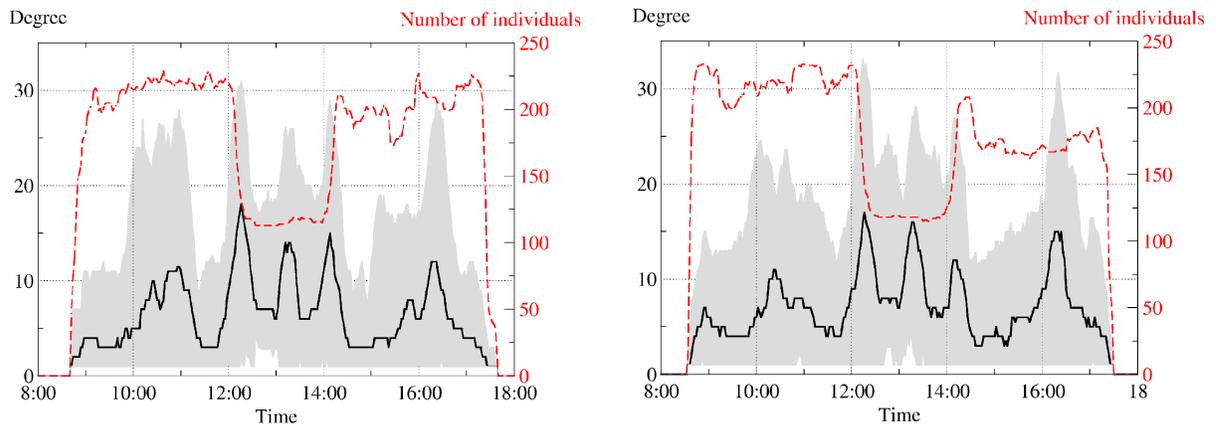

**Figure 6.** Degree of individuals in the contact networks aggregated over sliding time windows of 20 minutes during the first day (left) and the second day (right) of data collection. The median value is represented with a black line, the 95 % confidence interval is shown in gray and the number of individuals over which the statistics are calculated is shown in red dashes. Breaks and beginning and end of lunch are characterized by a sudden increase of the degree, showing the occurrence of large numbers of contact events.

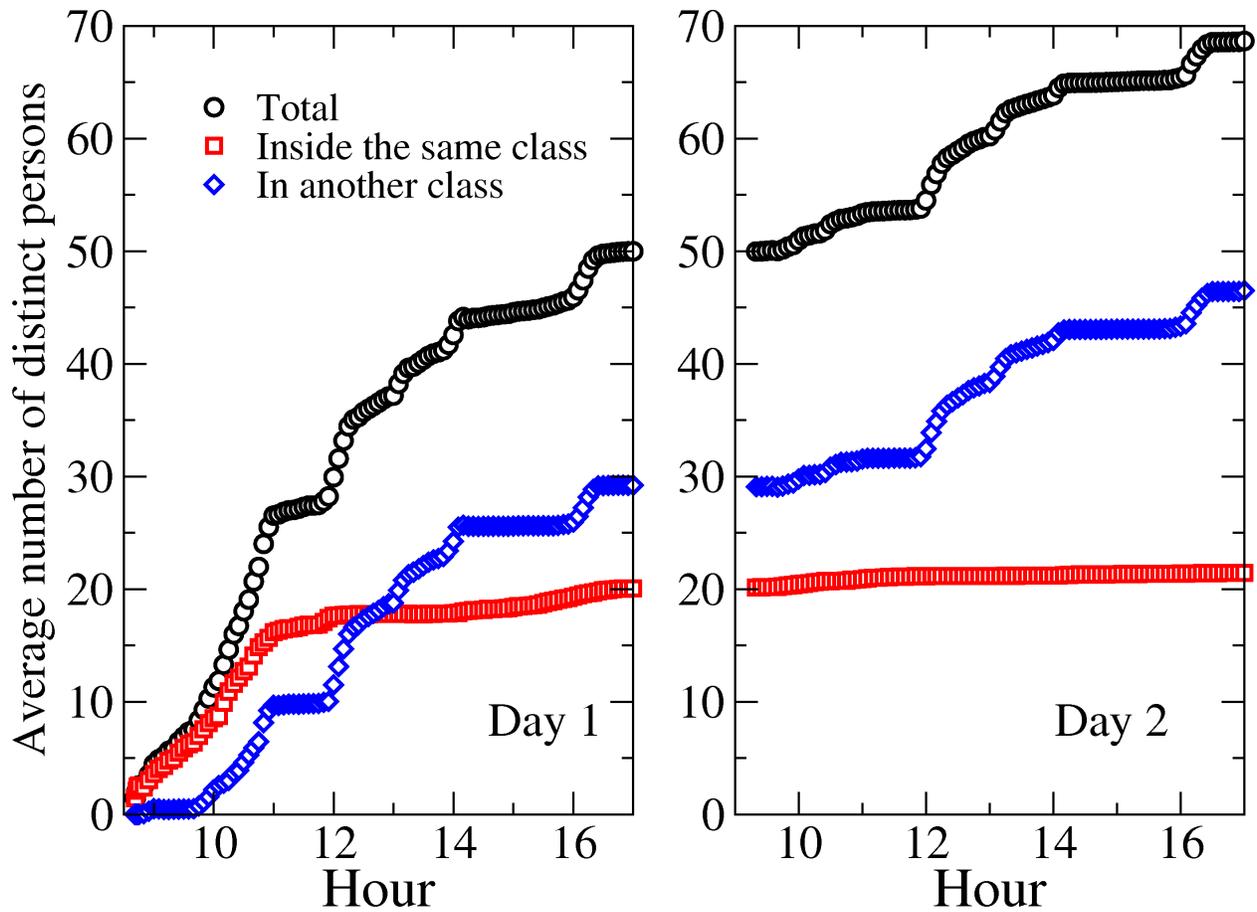

**Figure 7.** Time evolution of the average number of distinct children with whom a child has been in contact during the study. The average total number is displayed in black, the average number of children of the same class in red, and the average number of children of other classes in blue.

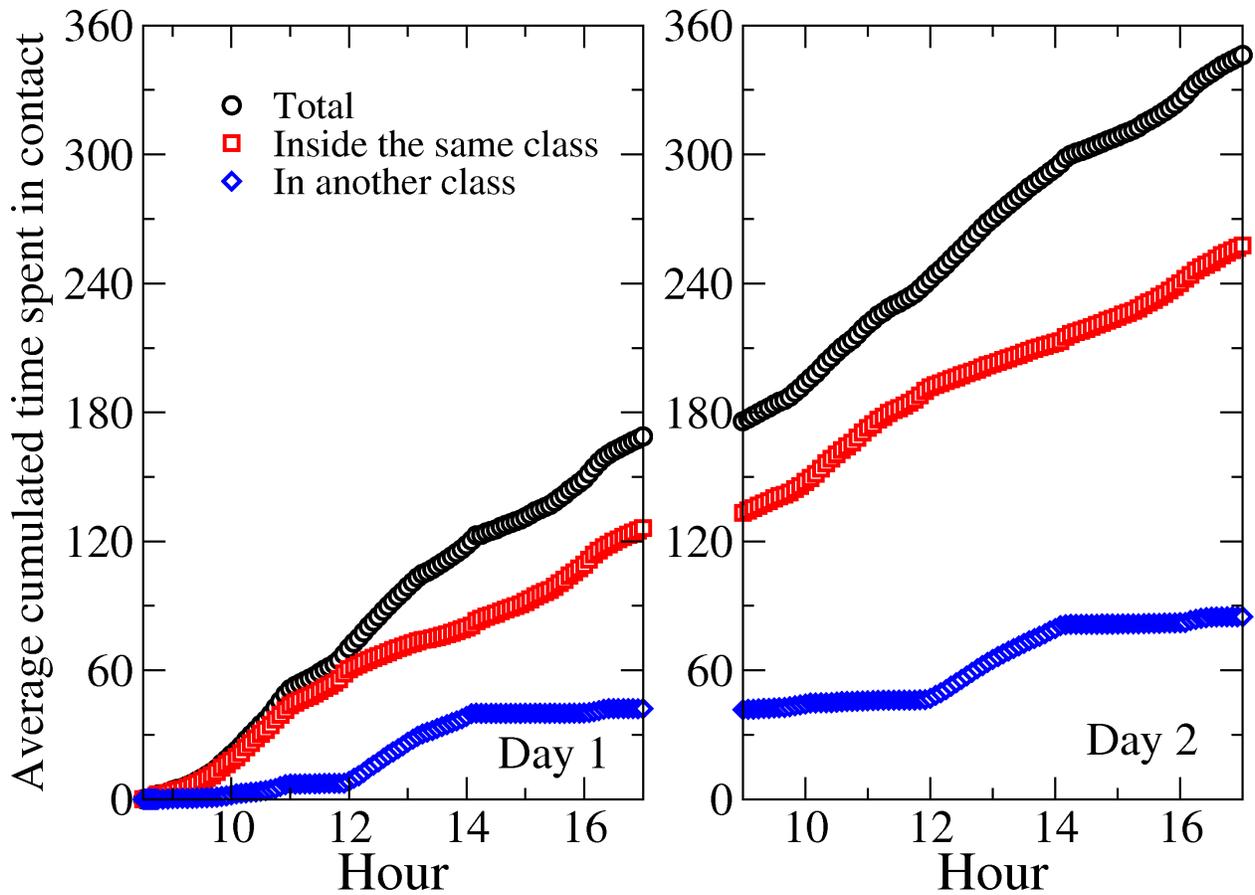

**Figure 8.** Time evolution of the average cumulated time spent by a child in contact with other children during the study. The average total time is displayed in black, the average time spent with children of the same class in red, and the average time spent with children of other classes in blue.

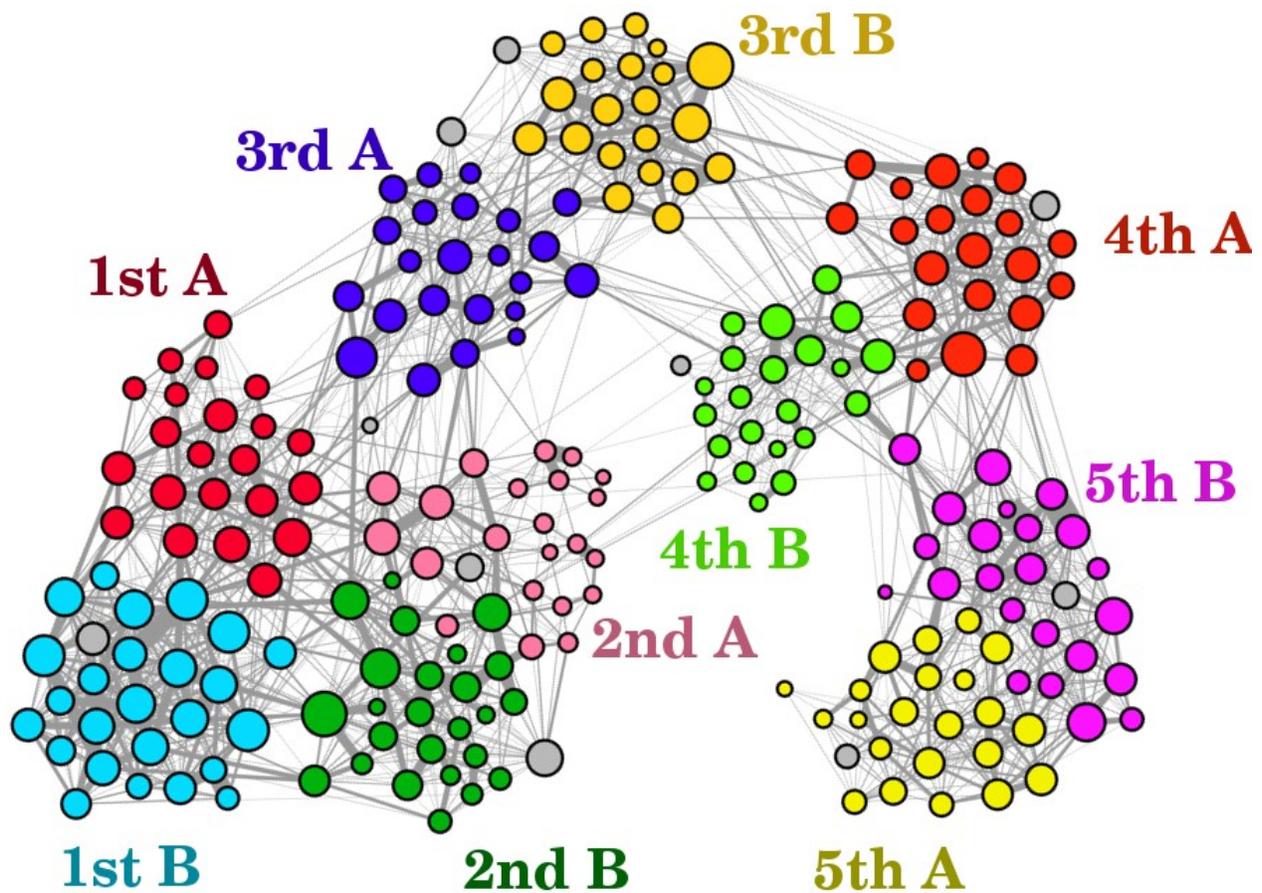

**Figure 9.** Network of contacts aggregated over the first day. Edges between individuals having interacted less than 2 minutes have been removed, thus keeping only the strongest links. The width of links corresponds to the cumulative duration of contacts, and nodes with higher number of edges have larger size. Colors correspond to classes, teachers are shown in grey. Figure created using the Gephi software, http://www.gephi.org.

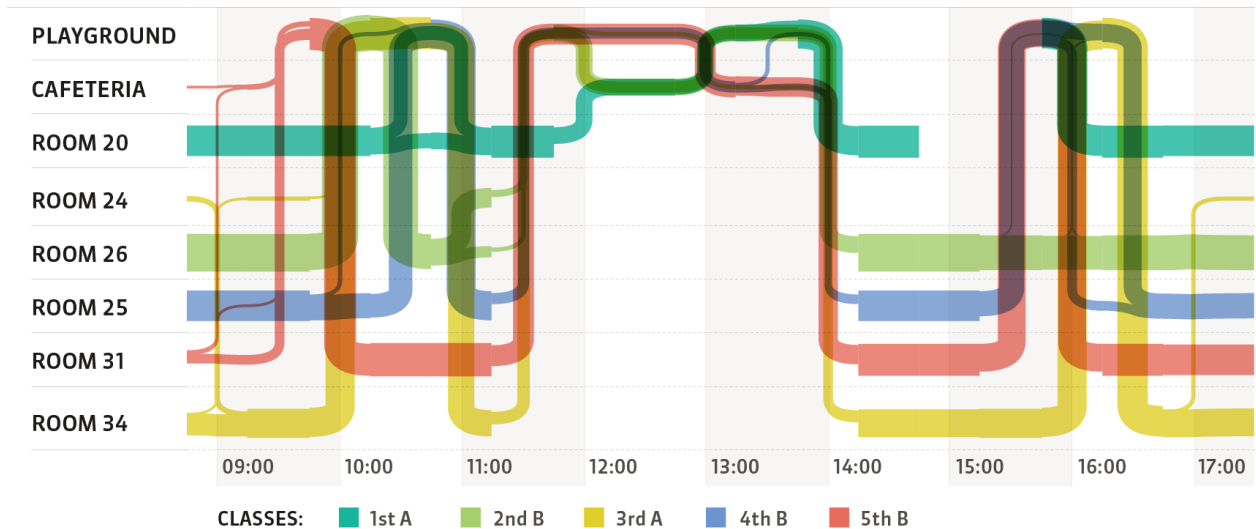

**Figure 10.** Approximate spatiotemporal trajectories of some classes. Each row corresponds to a particular place in the school (classroom, canteen, courtyard) where a RFID reader was situated, and each colored line corresponds to the spatio-temporal trajectory of the children of a class (only 5 classes are shown for clarity). Line widths correspond to the number of children whose approximate position correspond to the row area. A line can become thinner if children leave the school (for instance during the lunch break, to have lunch at home) or divide itself into two thinner lines if two groups of children of the same class follow distinct paths in the school   The trajectories highlight how mixing between classes, shown by the fact that the colored lines overlap, occurs during the breaks and is strongly constrained by the school schedule.

**Table 1.** Rate of participation for the school classes.

| Grade | Class name | Numbers of children or teachers | Number of participating children or teachers | | Participating rate (%) | |
|---|---|---|---|---|---|---|
| | | | Day 1 | Day 2 | Day 1 | Day 2 |
| 1 | Class 1A | 24 | 22 | 23 | 91.7 | 95.8 |
| | Class 1B | 25 | 25 | 25 | 100 | 100 |
| 2 | Class 2A | 25 | 22 | 23 | 88.0 | 92.0 |
| | Class 2B | 26 | 25 | 26 | 96.2 | 100 |
| 3 | Class 3A | 24 | 23 | 23 | 95.8 | 95.8 |
| | Class 3B | 22 | 21 | 21 | 95.5 | 95.5 |
| 4 | Class 4A | 23 | 21 | 21 | 91.3 | 91.3 |
| | Class 4B | 24 | 22 | 22 | 91.7 | 91.7 |
| 5 | Class 5A | 24 | 22 | 21 | 91.7 | 87.5 |
| | Class 5B | 24 | 23 | 23 | 95.8 | 95.8 |
| - | Teachers | 10 | 10 | 10 | 100 | 100 |

**Table 2:** Contact matrices between classes. The matrix entry for row A and column B gives the total number of contacts $n_{AB}$ measured between all individuals of classes A and B over the two days of data collection.

|  | 1st A | 1st B | 2nd A | 2nd B | 3rd A | 3rd B | 4th A | 4th B | 5th A | 5th B | teachers |
|---|---|---|---|---|---|---|---|---|---|---|---|
| 1st A | 4505 | 1051 | 594 | 625 | 560 | 286 | 83 | 160 | 57 | 105 | 149 |
| 1st B | 1051 | 9756 | 502 | 632 | 269 | 207 | 551 | 161 | 448 | 386 | 1084 |
| 2nd A | 594 | 502 | 5401 | 1583 | 657 | 360 | 77 | 56 | 76 | 30 | 586 |
| 2nd B | 625 | 632 | 1583 | 6270 | 712 | 373 | 119 | 36 | 41 | 54 | 508 |
| 3rd A | 560 | 269 | 657 | 712 | 5537 | 2076 | 77 | 163 | 109 | 82 | 414 |
| 3rd B | 286 | 207 | 360 | 373 | 2076 | 5926 | 248 | 193 | 154 | 219 | 282 |
| 4th A | 83 | 551 | 77 | 119 | 77 | 248 | 4496 | 828 | 351 | 745 | 382 |
| 4th B | 160 | 161 | 56 | 36 | 163 | 193 | 828 | 2843 | 119 | 346 | 168 |
| 5th A | 57 | 448 | 76 | 41 | 109 | 154 | 351 | 119 | 4913 | 1968 | 372 |
| 5th B | 105 | 386 | 30 | 54 | 82 | 219 | 745 | 119 | 1968 | 5025 | 273 |
| teachers | 149 | 1084 | 586 | 508 | 414 | 282 | 382 | 168 | 372 | 273 | 101 |

**Table 3:** Contact matrices between classes. The matrix entry for row A and column B gives the cumulated duration $w_{AB}$ (in minutes) of the contacts measured between all individuals of classes A and B over the two days of data collection.

|  | 1st A | 1st B | 2nd A | 2nd B | 3rd A | 3rd B | 4th A | 4th B | 5th A | 5th B | teachers |
|---|---|---|---|---|---|---|---|---|---|---|---|
| 1st A | 2242.3 | 582.7 | 315.3 | 340.0 | 260.7 | 126.3 | 30.3 | 61.3 | 20.0 | 37.0 | 61.7 |
| 1st B | 582.7 | 5611.0 | 234.3 | 367.0 | 119.3 | 83.7 | 271.0 | 84.3 | 197.0 | 169.0 | 459.7 |
| 2nd A | 315.3 | 234.3 | 3055.3 | 1068.3 | 339.3 | 219.0 | 30.0 | 19.3 | 25.66 | 11.7 | 331.7 |
| 2nd B | 340.0 | 367.0 | 1068.3 | 3723.0 | 365.7 | 179.7 | 53.3 | 16.0 | 14.66 | 20.0 | 247.3 |
| 3rd A | 260.67 | 119.3 | 339.3 | 365.7 | 2839.7 | 1105.3 | 29.7 | 75.3 | 40.33 | 30.0 | 201.7 |
| 3rd B | 126.3 | 83.67 | 219.0 | 179.7 | 1105.3 | 3436.3 | 117.7 | 85.7 | 56.0 | 85.0 | 147.3 |
| 4th A | 30.3 | 271.0 | 30.0 | 53.3 | 29.67 | 117.7 | 2421.7 | 439.3 | 163.0 | 373.0 | 179.7 |
| 4th B | 61.3 | 84.3 | 19.3 | 16.0 | 75.3 | 85.7 | 439.3 | 1600.0 | 46.0 | 207.3 | 68.3 |
| 5th A | 20.0 | 197.0 | 25.7 | 14.7 | 40.3 | 56.0 | 163.0 | 46.0 | 2671.0 | 966.7 | 188.3 |
| 5th B | 37.0 | 169.0 | 11.7 | 20.0 | 30.0 | 85.0 | 163.0 | 207.3 | 966.66 | 2752.7 | 134.7 |
| teachers | 61.67 | 459.67 | 331.7 | 247.3 | 201.67 | 147.3 | 179.7 | 68.3 | 188.33 | 134.7 | 65.0 |

**Table 4.** Exposure matrix between grades. The cell of row A and column B of the matrix gives the average number (and the duration in minutes, between parenthesis) of contacts involving an individual of grade A with any individual of grade B, per day.

|  | 1st grade | 2d grade | 3d grade | 4th grade | 5th grade | Teachers |
|---|---|---|---|---|---|---|
| **1st grade** | 322.6 (177.7) | 24.8 (13.3) | 13.9 (6.2) | 10.0 (4.7) | 10.4 (4.4) | 13.0 (5.5) |
| **2d grade** | 24.6 (13.2) | 274.8 (162.7) | 21.7 (11.4) | 3.0 (1.2) | 2.1 (0.7) | 11.4 (6.0) |
| **3d grade** | 15.0 (6.7) | 23.9 (12.5) | 307.7 (167.8) | 7.7 (3.5) | 6.4 (2.4) | 7.9 (4.0) |
| **4th grade** | 11.1 (5.2) | 3.3 (1.4) | 7.9 (3.6) | 189.9 (103.7) | 18.2 (9.2) | 6.4 (2.9) |
| **5th grade** | 11.3 (4.8) | 2.3 (0.8) | 6.3 (2.4) | 17.5 (8.9) | 269.3 (148.6) | 7.3 (3.6) |
| **Teachers** | 61.7 (26.1) | 54.8 (29.0) | 34.8 (17.5) | 27.5 (12.4) | 32.4 (16.2) | 10.5 (6.8) |

**Table 5.** Comparison of some characteristics of the networks of day 1 and 2.

|  | Day 1 | Day 2 |
|---|---|---|
| Number of individuals | 236 | 238 |
| Average number of contacts of an individual ($CV^2$) | 317 (0.22) | 338 (0.27) |
| Average total time in contact of an individual, in minutes ($CV^2$) | 172 (0.25) | 183 (0.33) |
| Average number of distinct persons contacted ($CV^2$) | 50 (0.14) | 46.5 (0.18) |
| Average cumulated time spent in contact by two persons, in seconds ($CV^2$) | 207 (5.4) | 236 (4.7) |
| Average duration of a contact, in seconds ($CV^2$) | 32.6 (1.2) | 32.6 (1.1) |
| Average clustering coefficient | 0.5 | 0.56 |

**Table 6. A.** Average number of children with whom a child was in contact, computed over one day, and average total time spent daily in contact with other children. In both cases, a filtering that retains only the contacts with duration at least equal to T is applied (T=0 or 20s corresponds to taking all contacts into account, given the available 20s time resolution). **B.** Average number of children with whom a child was in contact, computed over one day, and average total time spent daily in contact with other children. For both quantities, a filtering procedure is applied that retains only the links between children who have spent an amount of time at least equal to W in face-to-face proximity (W=0 or 20s corresponds to taking all links into account).

| A. Filtering procedure: only contacts of duration at least T | Average daily number of distinct other children in contact | Average daily cumulated duration of contacts with other children, in minutes |
|---|---|---|
| T=0 | 47.4 | 176 |
| T=40s | 20.8 | 100 |
| T=1 mn | 11.8 | 65 |
| T=2 mn | 4..1 | 28 |
| T=3 mn | 2.2 | 19 |
| | | |
| B. Filtering procedure: only cumulated contacts at least W | Average daily number of distinct other children in contact | Average daily cumulated duration of contacts with other children, in minutes |
| W=0 | 47.4 | 176 |
| W=1 mn | 21.4 | 163 |
| W=2 mn | 15.2 | 153 |
| W=5 mn | 8.1 | 129 |
| W=7 mn | 6.1 | 117 |
| W=10 mn | 4.3 | 102 |
| W=12 mn | 3.5 | 93 |
| W=15 mn | 2.7 | 81 |

**Table S1**

| Reference | Setting | Contact definition | Results |
|---|---|---|---|
| Mikolajczyk *et al*. [4] | Survey in a primary school; 6-10 years-old children. | A person with whom the child spoke or played with in a day | 25.1 contacts per day per child - |
| Wallinga *et al*. [7] | General population survey, divided into age classes: 1–5, 6–12, 13–19, 20–39, 40–59 and ≥60 years-old persons | Number of different conversation partners the participant encountered during a typical week by age classes. | 23.77 conversations per week (3.40 per day) held with different persons for 6-12 years-old children with other 6-12 years-old children. |
| Glass *et al*. [5] | High, middle and elementary schools surveys. Age classes: 10-12, 12-13, 14-15, 15-16, 16-17 and 17-18 years-old persons. | "An interaction with another person during which influenza could be passed. These must be within 3 feet and for a recognizable length of time" | 4.43 contacts per day for 10-12 years-old child with other 10-12 years-old children About 1 hour per day between 10-12 years-old children |
| Zagheni *et al*. [6] | School. 5-9, 10-14 and 15-19 years-old persons. | Estimation through time-use data, under the assumption of proportionate time mixing, of the co-presence of people in the same location. | At school, 98 min (1 hour 38 min) per day between 5-9 years-old children and 113 min (1 hour 53 min) per day between 10-14 years-old children. |
| Del Valle *et al*. [21] | General population, divided into age classes: 0-4, 5-12, 13-19, 20-29, 30-39, 40-49, 50-59, 60-69 and 70-90 years old. Data are obtained from the EpiSimS agent-based simulation of an entire city, based on US census statistics. | Co-presence in the same sub-location. The duration is defined as the total length that two people spent together in the same sub-location. The durations of multiple encounters between two persons are added up and the total aggregated length gives the final contact duration. | 3 hour 47 min between children at school (not detailed for age groups). |
| Mossong *et al*. [8] | General population in 8 European countries, divided into age-classes: 0-4, 5-9, 10-14, 15-19, 20-29, 30-39, 40-49, 50-59, 60-69 and ≥70 years-old persons | Either skin-to-skin contact such as a kiss or handshake (a physical contact), or a two-way conversation with three or more words in the physical presence of another person but no skin-to-skin contact (a nonphysical contact) | Average number on all reported contact persons (physical and non-physical contacts) per day per person: From 2.25 to 11.88 between 5-9 years old children, depending on the country From 3.58 to 14.56 between 10-14 years-old children depending on the country |
| Salathé *et al*. [3] | US high school: students, teachers and staff. | Electronic devices (motes). A close proximity record CPR | On average 1900 contacts per student per day, lasting on average about 1 |

| Reference | Setting | Contact definition | Results |
|---|---|---|---|
| | | represents one close (≤3 meters) proximity detection event between two electronic devices. A contact is defined as a continuous sequence of CPR between two motes. | minute. Broad distributions of the duration of contacts and of the cumulative time spent in proximity by two individuals. Each individual has contact with an average of 300 distinct other individuals. |
| Stehle *et al.* (present study) | Primary school: 6-12 years-old children | RFID devices that exchange radio packets only when the individuals wearing them face each other at close range (about 1 to 1.5 m) | On average 323 contacts per child per day, lasting on average 33 seconds, with on average 47 other distinct individuals. Cumulated contact time of each individual of 176 min (2 hours 56 min) per day on average. Broad distributions of the duration of contacts and of the cumulative time spent in proximity by two individuals. |